\documentclass[fleqn,twocolumn,pre]{revtex4-2}

\usepackage{amssymb,amsmath,pifont,graphicx,color}

\def\bq{\begin{equation}}
\def\ee{\end{equation}}
\def\lt{\left}
\def\rt{\right}
\def\md{\,\mbox{d}}
\def\f{\frac}

\def\de{\delta}
\def\D{\Delta}

\def\GP{G_{\mbox{\scriptsize P}}}

\def\sP{\sigma_{\mbox{\tiny P}}}

\newcommand\lae[1]{\label{#1}}

\begin{document}

\title{New concept to measure the Poiseuille coefficient of rarefied gas flow driven by a large pressure drop}
\author{Felix Sharipov$^a$}
\email{sharipov@ufpr.br;sharipov@fisica.ufpr.br}
\affiliation{organization={Departamento de F\'\i sica, Universidade Federal do Paran\'a}, address={81531-980 Curitiba, Brazil}}

\author{Irina Graur$^b$}
\email{irina.martin@univ-amu.fr}
\affiliation{IUSTI, Aix-Marseille Univ, CNRS, UMR 7343, France}

\begin{abstract}
The Poiseuille coefficient, which relates the mass flow rate through a long capillary to the local pressure gradient, is an important characteristic in designing various technological processes that include vacuum systems as part. This quantity is usually calculated by solving numerically the linearized Boltzmann equation assuming a  small pressure gradient and, therefore, the mass flow rate is  small  too.  As a result, to extract accurately this coefficient from experimental measurements, the mass flow rate must remain small. The present paper proposes a novel approach to extract the Poiseuille coefficient from a measurement of mass flow rate driven by a large pressure drop. Under these conditions, the flow rate increases significantly  and, hence, the experimental error is reduced. The proposed method will help experimentalists in designing measurement devices to  determine the Poiseuille coefficient with a higher accuracy.

\end{abstract}

\maketitle

\section{Introduction}

Over the past decades, research in rarefied gas dynamics has advanced significantly, driven by the growing  demand of vacuum technology, aerospace research, miniaturization of technological processes, etc.  Most progress has been made in theoretical approaches, including the development of new numerical methods for solving of kinetic equations \cite{CerB1,Sha02B}, improvements in the direct simulation Monte Carlo technique \cite{Bir02}, and implementation of {\it ab initio} calculations \cite{Sha145,Dod01}. However, experimental data on several aspects of rarefied gas dynamics   remain scarce, and their experimental uncertainty is  relatively high. Therefore, further improvement of experimental methods is  essential to obtain more reliable data.

The Poiseuille coefficient, which relates the mass flow rate through a long capillary to the local pressure gradient, is widely used to model various processes in vacuum systems. For certain simple capillary cross-sections, such as cylindrical \cite{Cer16,Loy20,Sha45}, rectangular \cite{Loy20,Sha32}, elliptical \cite{Sha67}, and annular \cite{Bas02,Bre04} tubes, this coefficient has been calculated across all flow regimes, namely: free-molecular, transitional, and viscous. All these results have been obtained from the linearized Boltzmann equation or its kinetic models \cite{CerB1,Sha02B}, which assume a low pressure gradient and consequently, a low mass flow rate. Despite this assumption, the Poiseuille coefficient is commonly used to calculate the mass flow rate through a long capillary driven by a high pressure drop \cite{Sha15}. Additionally, gas flows through a capillary with a variable cross-section can also be  efficiently calculated via the Poiseuille coefficient \cite{Sha57,Gra11}.

In many vacuum systems, the flow regime is viscous at the capillary inlet  and free-molecular  at the outlet, see e.g. Refs. \cite{Sha72,Sha112}. A direct solution of such a scenario requires significant computational effort. However, knowledge of the Poiseuille coefficient  enables the calculation of both flow rate and pressure distribution along the capillary with modest computer resources. This is achieved using a general approach to transient flows in rarefied gases, as proposed in Ref. \cite{Sha103}, which is based on the Poiseuille coefficient.

Besides capillary shape, the Poiseuille coefficient is influenced by several other factors, including the rarefaction parameter (inverse Knudsen number), gas species, chemical composition in case of gaseous mixture, and accommodation coefficients that characterize gas-surface interaction, etc. As a result, theoretical predictions of the Poiseuille coefficient are not always directly applicable to practical situations. Moreover, accommodation coefficients are often unknown and their determination  through Molecular Dynamics simulations is complex \cite{Spi01} and frequently deviates from experimental assesses. Therefore, obtaining the Poiseuille coefficient experimentally is essential. Such experimental results could help validate theoretical models and extract accommodation coefficients for  specific gas-surface interactions.

The mass flow rate is usually measured via the pressure variation in the inlet volume connected to a capillary.
The experimental data reported by Porodnov et al. \cite{Aki01,Por02,Sue02} are based on the assumption that the initial pressure difference between the capillary ends is small and decreases exponentially over time. In this case, the resulting flow rate is small leading to a large experimental error. When the initial pressure drop is large, the authors of Ref. \cite{Por04} did not extract the Poiseuille coefficient, but  instead  reported the ratio of the flow rate at an arbitrary Knudsen number to its value in the free-molecular regime. Similarly, the authors of Refs. \cite{Ewa01,Ewa02,Nac01,Per04,Roj02} considered a large initial pressure drop and reported the dimensional mass flow rate without extracting the Poiseuille coefficient.

Some studies extract the Poiseuille coefficient from experiments with a large initial pressure drop, using the average pressure gradient and rarefaction parameter based on the average pressure, see e.g. Refs.\cite{Had09,Sil03,Pit01,Var01}. However, this approximation introduces an error, that can be significant in the case of wide rectangular channel \cite{Sha32} and elliptical tube \cite{Sha67}. In fact, the Knudsen minimum for these geometrical configurations is deep that hinders  the use of average pressure to extract the Poiseuille coefficient.

The present work proposes  a new method to extract the Poiseuille coefficient from  measurements of pressure decay in a reservoir caused by gas flow through a capillary driven by large pressure drop.

\section{Formulation of the problem}

The Poiseuille coefficient $\GP$ relates the mass flow rate through a long capillary to the local pressure gradient as \cite{Sha02B}
\begin{equation}
\dot M=-\GP \f{A a}{v_m}\f{\md p}{\md x},
\lae{AG}\end{equation}
where $A$ is the cross-section area of the capillary, $a$ is its characteristic transversal size, $v_m=\sqrt{2R_gT/{\cal M}}$ is the most probable molecular speed, $R_g=8.314$ J/(K mol) is the universal gas constant, $T$ is the gas temperature, ${\cal M}$ is the molar weight of the gas, $p$ is the gas pressure, and $x$ is the longitudinal coordinate directed from the inlet to the outlet of a capillary. The main parameter determining the Poiseuille coefficient is the rarefaction parameter
\begin{equation}
\de=ap/(\mu v_m),
\lae{AH}\end{equation}
with $\mu$ being the dynamic viscosity. Thus,  large values of $\de$ indicate the viscous regime, while the limit $\de\to0$ corresponds to the free-molecular regime.

In many application, the throughput $q$ is used to characterize the flow rate, which is related to the mass flow rate $\dot M$ as
\begin{equation}
q=\f12 v_M^2 \dot M.
\lae{AL}\end{equation}
Consider a long capillary connecting two vacuum chambers as shown in Figure \ref{figA}. The pressure in the inlet chamber $p_{in}$ is higher than that in the outlet chamber  ($p_{in}>p_{out}$), so that the gas flows to the right according to Figure \ref{figA}.
We assume that the gas flow in the capillary is stationary meaning that the mass flow rate remains constant along the capillary even though $p_{in}$ and $p_{out}$  vary. The assumption holds when the volumes $V_{in}$ and $V_{out}$ are significantly larger than the capillary volume ($V_c=AL$), as stated in Refs. \cite{Sha103,Sha99}.
When the pressures $p_{in}$ and $p_{out}$ are known, the steady throughput is easily calculated by integrating the function $\GP(\de)$ \cite{Sha15}
\bq
q=\f{A a v_m}{2L} \f{p_{in}-p_{out}}{\de_{in}-\de_{out}}\int_{\de_{out}}^{\de_{in}} \GP(\de)\md \de,
\lae{AD}\ee
where $\de_{in}$ and $\de_{out}$ are calculated by Eq.(\ref{AH}) for $p_{in}$ and $p_{out}$, respectively. This expression shows how the Poiseuille coefficient obtained from  the linearized Boltzmann equation is applied to calculate the throughput of gases driven by a large pressure drop. Knowledge of $\GP$ also allows to calculate a variation of the pressure $p_{in}$ and $p_{out}$  over time \cite{Sha103}. However, the inverse problem, i.e., calculation of the Poiseuille coefficient from the functions $p_{in}=p_{in}(t)$ and $p_{out}=p_{out}(t)$ is not trivial. Below, we derive the Poiseuille coefficient assuming that the functions  $p_{in}(t)$ and $p_{out}(t)$ are known.

\begin{figure}
  \centering
  \includegraphics[width=6cm]{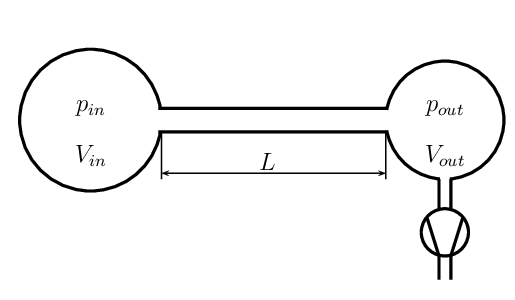}
  \caption{Scheme of the flow}\lae{figA}
\end{figure}

\section{Derivations}

We consider a setup where the inlet chamber is connected to the outlet chamber solely through the capillary, while the outlet chamber may either  be connected only to the capillary or also pumped.
In this case, the throughput is related to the variation of the inlet pressure by
\bq
q=-V_{in}\f{\md p_{in}}{\md t},
\lae{AC}\ee
where $V_{in}$ is the inlet chamber volume. A combination of this relation with Eq.(\ref{AD}) leads to
\bq
\tau\f{\md p_{in}}{\md t}=-\int_{p_{out}}^{p_{in}} \GP(\de)\md p,
\lae{AN}\ee
where $\tau$ is the characteristic time of variation of inlet pressure $p_{in}$
\bq
\tau=\f{2 LV_{in}}{A a v_m }.
\lae{AP}\ee
Calculating the derivative of both sides of Eq.(\ref{AN}), we have
\begin{equation}
\tau \f{\md^2 p_{in}}{\md t^2}=-\f{\md p_{in}}{\md t}\GP(\de_{in})
+\f{\md p_{out}}{\md t}\GP(\de_{out}) .
\lae{AK}\end{equation}
This procedure allows us to eliminate the integral of the function $\GP$, but expression (\ref{AK}) still contains two unknown values of $\GP$ function. The last term in Eq.(\ref{AK}) can be eliminated if the outlet pressure $p_{out}$ is constant. This condition can be reached, for example, if the vacuum pump connected to the outlet chamber has a high pumping speed keeping the pressure $p_{out}$ low enough. Then, the variation of $p_{out}$ will be significantly lower than that of $p_{in}$, i.e.,
\begin{equation}
\lt|\f{\md p_{out}}{\md t}\rt| \ll \lt|\f{\md p_{in}}{\md t}\rt|.
\lae{BX}\end{equation}
An additional factor  that can reduce the variation of $p_{out}$ is the outlet volume $V_{out}$. Since the characteristic time for the variation of $p_{out}$ is given by $\tau_{out}=2LV_{out}/(A a v_m)$, the condition $V_{out}\gg V_{in}$ will reinforce the inequality (\ref{BX}) and may even be the only factor to maintain $p_{out}$ constant.

 Assuming (\ref{BX}), Eq. (\ref{AK}) is reduce to
\begin{equation}
\GP(\de_{in})=-\tau \f{\md^2 p_{in}}{\md t^2}\lt(\f{\md p_{in}}{\md t}\rt)^{-1}.
\lae{AM}\end{equation}

Thus, if all the quantities required to calculate the characteristic time $\tau$ given by (\ref{AP}) are known and $p_{in}$ is measured during an experiment as a function of time, the Poiseuille coefficient  can be calculated from Eq.(\ref{AM}) by applying an interpolation procedure to the measured inlet pressure. Since the experimental data for $p_{in}$ are provided at discrete time points, interpolation of these data is necessary to calculate the first and second derivatives. One possible interpolation  formula is as follows
\[
p_{in}(t)=p_{in}(t_0)\exp\lt[C_1(t-t_0)\rt.
\]
\begin{equation}\hskip1cm\lt.
+C_2(t-t_0)^2+C_3(t-t_0)^3\rt],
\lae{AV}\end{equation}
where the coefficients $C_1$, $C_2$, and $C_3$ are calculated by the least squares. The interpolation is performed in a certain interval  $[t_0-\D t,t_0+\D t]$, where $\D t$ should be large to contain a sufficiently large number of experimental points. At the same time, it should cover an interval with a moderate variation of $p_{in}(t)$. Once the interpolating coefficients are known, the derivative are computed as
\begin{equation}
\lt.\f{\md p_{in}}{\md t}\rt|_{t=t_0}=C_1p_{in}(t_0),
\lae{AX}\end{equation}
\begin{equation}
\lt.\f{\md^2 p_{in}}{\md t^2}\rt|_{t=t_0}=\lt(C_1^2+2C_2\rt)p_{in}(t_0).
\lae{AY}\end{equation}
Substituting (\ref{AX}) and (\ref{AY}) into (\ref{AM}), we have
\begin{equation}
\GP(\de_{in})=-\tau \lt(C_1+2\f{C_2}{C_1}\rt).
\lae{AZ}\end{equation}
The experimental error of the Poiseuille coefficient $\GP$ calculated from Eq.(\ref{AZ}), includes the uncertainties of the parameters determining the characteristic time $\tau$ given by Eq.(\ref{AP}), namely, the length of the capillary $L$, its cross-section  $A$, and the volume of the inlet chamber $V_{in}$.

In the viscous and slip regimes the Poiseuille coefficient is usually known analytically and has the form
\begin{equation}
\GP=H\de+S\sP,
\lae{AQ}\end{equation}
where the viscous limit solution $H$ and slip correction $S$ depend on the cross-section shape, but they are independent from the gas species. For instance, $H=1/4$ and $S=1$ for a cylindrical tube. In the case of rectangular channel, the expressions of $H$ and $S$ are given in Chap.13 of the book \cite{Sha02B}. Their expressions for an elliptical and annular tubes are given in Refs.\cite{Sha67} and \cite{Bre04}, respectively. The slip coefficient $\sP$ depends on the tangential momentum accommodation coefficient \cite{Sha44}, which characterizes gas-surface interaction. If we assume that this interaction is diffuse, which is the case for most of gases, so that the value $\sP=1.02$ can be used \cite{Sha02B}. Thus, the expression of $\GP$ in the viscous regime (\ref{AQ}) can be used to extract the characteristic time $\tau$  from the measurements. Then, the coefficient $\GP$ is calculated from Eq. (\ref{AZ}) using the measured inlet pressure over the whole range of the gas rarefaction.

\section{Conclusion}

It is shown that a measurement of the pressure variation at the inlet of a long capillary can be used to extract the Poiseuille coefficient with high accuracy. In turn, this coefficient enables the calculation of the transient flow rate through the capillary and the pressure distribution inside it for an arbitrary pressure drop. More reliable experimental data on the Poiseuille coefficient will allow to determine the accommodation coefficients more accurately.

\section*{CRediT authorship contribution statement}

{\bf Felix Sharipov}: Conceptualization, Methodology, Formal analysis, Writing – original draft, Writing – review \& editing.

{\bf Irina Graur}: Conceptualization, Formal analysis, Funding acquisition, Writing – review \& editing.

\section*{Declaration of competing interest}
The authors declare that they have no known competing financial interests or personal relationships that could have appeared to influence the work reported in this paper.

\section*{Acknowledgments}
F.S. thanks CNPq (Brazil) for supporting his research (grant No 303429/2022-4). The Aix-Marseille university is acknowledged for the support of the visit of F.S.

\bibliographystyle{elsarticle-num}

\begin{thebibliography}{10}
\expandafter\ifx\csname url\endcsname\relax
  \def\url#1{\texttt{#1}}\fi
\expandafter\ifx\csname urlprefix\endcsname\relax\def\urlprefix{URL }\fi
\expandafter\ifx\csname href\endcsname\relax
  \def\href#1#2{#2} \def\path#1{#1}\fi

\bibitem{CerB1}
C.~Cercignani, The {B}oltzmann Equation and its Application, Springer-Verlag,
  New York, 1988.

\bibitem{Sha02B}
F.~Sharipov, Rarefied Gas Dynamics. Fundamentals for Research and Practice,
  Wiley-VCH, Berlin, 2016.
\newblock \href {https://doi.org/10.1002/9783527685523}
  {\path{doi:10.1002/9783527685523}}.

\bibitem{Bir02}
G.~A. Bird, Molecular Gas Dynamics and the Direct Simulation of Gas Flows,
  Oxford University Press, Oxford, 1994.
\newblock \href {https://doi.org/10.1093/oso/9780198561958.001.0001}
  {\path{doi:10.1093/oso/9780198561958.001.0001}}.

\bibitem{Sha145}
F.~Sharipov, Ab initio modelling of transport phenomena in multi-component
  mixtures of rarefied gases, Int. J. Heat Mass Transfer 220 (2024) 124906.
\newblock \href {https://doi.org/10.1016/j.ijheatmasstransfer.2023.124906}
  {\path{doi:10.1016/j.ijheatmasstransfer.2023.124906}}.

\bibitem{Dod01}
O.~I. Dodulad, Y.~Y. Kloss, D.~O. Savichkin, F.~G. Tcheremissine, Knudsen pumps
  modeling with {L}ennard-{J}ones and ab initio intermolecular potentials,
  Vacuum 109 (2014) 360--367.
\newblock \href {https://doi.org/10.1016/j.vacuum.2014.06.019}
  {\path{doi:10.1016/j.vacuum.2014.06.019}}.

\bibitem{Cer16}
C.~Cercignani, F.~Sernagiotto, Cylindrical {P}oiseuille flow of a rarefied gas,
  Phys. Fluids 9~(1) (1966) 40--44.
\newblock \href {https://doi.org/10.1063/1.1761530}
  {\path{doi:10.1063/1.1761530}}.

\bibitem{Loy20}
S.~K. Loyalka, S.~A. Hamoodi, Poiseuille flow of a rarefied gas in a
  cylindrical tube: Solution of linearized {B}oltzmann equation, Phys.Fluids A
  2~(11) (1990) 2061--2065, erratum. in Phys. Fluids A \textbf{ 3}, 2825
  (1991).
\newblock \href {https://doi.org/10.1063/1.857681}
  {\path{doi:10.1063/1.857681}}.

\bibitem{Sha45}
F.~Sharipov, Application of the {C}ercignani-{L}ampis scattering kernel to
  calculations of rarefied gas flows. {III}. {P}oiseuille flow and thermal
  creep through a long tube, Eur. J. Mech. B / Fluids 22 (2003) 145--154.
\newblock \href {https://doi.org/10.1016/S0997-7546(03)00018-9}
  {\path{doi:10.1016/S0997-7546(03)00018-9}}.

\bibitem{Sha32}
F.~Sharipov, Rarefied gas flow through a long rectangular channel, J. Vac. Sci.
  Technol. A 17~(5) (1999) 3062--3066.
\newblock \href {https://doi.org/10.1116/1.582006}
  {\path{doi:10.1116/1.582006}}.

\bibitem{Sha67}
I.~Graur, F.~Sharipov, Gas flow through an elliptical tube over the whole range
  of the gas rarefaction, Eur. J. Mech. B / Fluids 27~(3) (2007) 335--345.
\newblock \href {https://doi.org/10.1016/j.euromechflu.2007.07.003}
  {\path{doi:10.1016/j.euromechflu.2007.07.003}}.

\bibitem{Bas02}
P.~Bassanini, C.~Cercignani, F.~Sernagiotto, Flow of a rarefied gas in a tube
  of annular section, Phys. Fluids 9~(6) (1966) 1174--1178.
\newblock \href {https://doi.org/10.1063/1.1761817}
  {\path{doi:10.1063/1.1761817}}.

\bibitem{Bre04}
G.~Breyiannis, S.~Varoutis, D.~Valougeorgis, Rarefied gas flow in concentric
  annular tube: {E}stimation of {P}oiseuille number and the exact hydraulic
  diameter, Eur. J. Mech. B/Fluids 27 (2008) 609--622.
\newblock \href {https://doi.org/10.1016/j.euromechflu.2007.10.002}
  {\path{doi:10.1016/j.euromechflu.2007.10.002}}.

\bibitem{Sha15}
F.~Sharipov, V.~Seleznev, Rarefied gas flow through a long tube at any pressure
  ratio, J. Vac. Sci. Technol. A 12~(5) (1994) 2933--2935.
\newblock \href {https://doi.org/10.1116/1.578969}
  {\path{doi:10.1116/1.578969}}.

\bibitem{Sha57}
F.~Sharipov, G.~Bertoldo, Rarefied gas flow through a long tube of variable
  radius, J. Vac. Sci. Technol. A 23~(3) (2005) 531--533.
\newblock \href {https://doi.org/10.1116/1.1897703}
  {\path{doi:10.1116/1.1897703}}.

\bibitem{Gra11}
I.~Graur, M.~T. Ho, Rarefied gas flow through a long rectangular channel of
  variable cross section, Vacuum 101 (2014) 328--332.
\newblock \href {https://doi.org/10.1016/j.vacuum.2013.07.047}
  {\path{doi:10.1016/j.vacuum.2013.07.047}}.

\bibitem{Sha72}
O.~B. Malyshev, C.~Day, X.~Luo, F.~Sharipov, Tritium gas flow dynamics through
  the source and transport system of the {KATRIN} experiment, J. Vac Sci.
  Technol. A 27~(1) (2009) 73--81.
\newblock \href {https://doi.org/10.1116/1.3039679}
  {\path{doi:10.1116/1.3039679}}.

\bibitem{Sha112}
F.~Sharipov, Y.~Yang, J.~E. Ricker, J.~H. Hendricks, Primary pressure standard
  based on piston-cylinder assemblies. calculation of effective cross sectional
  area based on rarefied gas dynamics, Metrologia 53~(5) (2016) 1177--1184.
\newblock \href {https://doi.org/10.1088/0026-1394/53/5/1177}
  {\path{doi:10.1088/0026-1394/53/5/1177}}.

\bibitem{Sha103}
F.~Sharipov, I.~Graur, General approach to transient flows of rarefied gases
  through long capillaries, Vacuum 100 (2014) 22--25.
\newblock \href {https://doi.org/10.1016/j.vacuum.2013.07.017}
  {\path{doi:10.1016/j.vacuum.2013.07.017}}.

\bibitem{Spi01}
P.~Spijker, A.~J. Markvoort, S.~V. Nedea, P.~A.~J. Hilbers, Computation of
  accommodation coefficients and the use of velocity correlation profiles in
  molecular dynamics simulations, Phys. Rev. E 81 (2010) 011203.
\newblock \href {https://doi.org/10.1103/PhysRevE.81.011203}
  {\path{doi:10.1103/PhysRevE.81.011203}}.

\bibitem{Aki01}
V.~D. Akinshin, S.~F. Borisov, B.~T. Porodnov, P.~E. Suetin, Flow of rarefied
  gases in a capillary screen at different temperatures, Journal of Applied
  Mechanics and Technical Physics 15~(2) (1974) 184--186.
\newblock \href {https://doi.org/10.1007/BF00850656}
  {\path{doi:10.1007/BF00850656}}.

\bibitem{Por02}
B.~T. Porodnov, P.~E. Suetin, S.~F. Borisov, V.~D. Akinshin, Experimental
  investigation of rarefied gas flow in different channels, J. Fluid Mech.
  64~(3) (1974) 417--437.
\newblock \href {https://doi.org/10.1017/S0022112074002485}
  {\path{doi:10.1017/S0022112074002485}}.

\bibitem{Sue02}
P.~E. Suetin, B.~T. Porodnov, V.~G. Chernyak, S.~F. Borisov, Poiseuille flow at
  arbitrary {K}nudsen number and tangential momentum accommodation, J. Fluid
  Mech. 60~(3) (1973) 581--592.
\newblock \href {https://doi.org/10.3390/mi11030319}
  {\path{doi:10.3390/mi11030319}}.

\bibitem{Por04}
B.~T. Porodnov, A.~G. Flyagin, Experimental investigation of the escape of
  helium, neon, and argon in a vacuum through a long single capillary at the
  temperatures 295 - 490k, Journal of Applied Mechanics and Technical Physics
  19~(4) (1978) 431--436.
\newblock \href {https://doi.org/10.1007/BF00859387}
  {\path{doi:10.1007/BF00859387}}.

\bibitem{Ewa01}
T.~Ewart, P.~Perrier, I.~Graur, J.~G. M\'eolans, Mass flow rate measurements in
  gas micro flows, Experiments in Fluids 41~(3) (2006) 487--498.
\newblock \href {https://doi.org/10.1007/s00348-006-0176-z}
  {\path{doi:10.1007/s00348-006-0176-z}}.

\bibitem{Ewa02}
T.~Ewart, P.~Perrier, I.~A. Graur, J.~G. M\'eolans, Mass flow rate measurements
  in a microchannel, from hydrodynamic to near free molecular regimes, J. Fluid
  Mech. 584 (2007) 337--356.
\newblock \href {https://doi.org/10.1017/S0022112007006374}
  {\path{doi:10.1017/S0022112007006374}}.

\bibitem{Nac01}
M.~H. Nacer, I.~Graur, P.~Perrier, Mass flow measurement through rectangular
  microchannel from hydrodynamic to near free molecular regimes, Houille
  Blanche-Rev. Int.~(4) (2011) 49--54.
\newblock \href {https://doi.org/10.1051/lhb/2011040}
  {\path{doi:10.1051/lhb/2011040}}.

\bibitem{Per04}
P.~Perrier, I.~A. Graur, T.~Ewart, J.~G. Meolans, Mass flow rate measurements
  in microtubes: From hydrodynamic to near free molecular regime, Phys. Fluids
  23~(4) (2011) 042004.
\newblock \href {https://doi.org/10.1063/1.3562948}
  {\path{doi:10.1063/1.3562948}}.

\bibitem{Roj02}
M.~Rojas-C{\'a}rdenas, E.~Silva, M.~T. Ho, C.~J. Deschamps, I.~Graur,
  Time-dependent methodology for non-stationary mass flow rate measurements in
  a long micro-tube, Microfluidics and Nanofluidics 21~(5) (2017) 86.
\newblock \href {https://doi.org/10.1007/s10404-017-1920-9}
  {\path{doi:10.1007/s10404-017-1920-9}}.

\bibitem{Had09}
M.~Hadj~Nacer, I.~Graur, P.~Perrier, J.~G. M\'eolans, M.~W\"{u}est, Gas flow
  through microtubes with different surface coating, J. Vac. Sci. Technol. A
  32~(2) (2014) 021601.
\newblock \href {https://doi.org/10.1116/1.4828955}
  {\path{doi:10.1116/1.4828955}}.

\bibitem{Sil03}
E.~Silva, C.~Deschamps, M.~Rojas-Cardenas, C.~Barrot-Lattes, L.~Baldas,
  S.~Colin, A time-dependent method for the measurement of mass flow rate of
  gases in microchannels, Int. J. Heat Mass Transfer 120 (2018) 422--434.
\newblock \href {https://doi.org/10.1016/j.ijheatmasstransfer.2017.11.147}
  {\path{doi:10.1016/j.ijheatmasstransfer.2017.11.147}}.

\bibitem{Pit01}
J.~Pitakarnnop, S.~Varoutis, D.~Valougeorgis, S.~Geoffroy, L.~Baldas, S.~Colin,
  A novel experimental setup for gas microflows, Microfluidics and Nanofluidics
  8~(1) (2010) 57--72.
\newblock \href {https://doi.org/doi.org/10.1007/s10404-009-0447-0}
  {\path{doi:doi.org/10.1007/s10404-009-0447-0}}.

\bibitem{Var01}
S.~Varoutis, S.~Naris, V.~Hauer, C.~Day, Computational and experimental study
  of gas flows through long channels of various cross sections in the whole
  range of the {K}nudsen number, J. Vac. Sci. Technol. A 27~(1) (2009) 89--100.
\newblock \href {https://doi.org/10.1116/1.3043463}
  {\path{doi:10.1116/1.3043463}}.

\bibitem{Sha99}
F.~Sharipov, Transient flow of rarefied gas through a short tube, Vacuum 90
  (2013) 25--30.

\bibitem{Sha44}
F.~Sharipov, Application of the {C}ercignani-{L}ampis scattering kernel to
  calculations of rarefied gas flows. {II}. {S}lip and jump coefficients, Eur.
  J. Mech. B / Fluids 22 (2003) 133--143.
\newblock \href {https://doi.org/10.1016/S0997-7546(03)00017-7}
  {\path{doi:10.1016/S0997-7546(03)00017-7}}.

\end{thebibliography}

\end{document}